\documentclass[twocolumn,aps,prd,superscriptaddress,showpacs,floatfix]{revtex4}

\usepackage{latexsym}
\usepackage{amsmath}
\usepackage{amssymb}
\usepackage{latexsym}
\usepackage{color}
\usepackage{graphicx}
\usepackage{cancel}
\usepackage{bbm}
\usepackage{bm}
\usepackage{maybemath}

\newcommand{\dd}{\mathrm{d}}
\newcommand{\ee}{\mathrm{e}}
\newcommand{\ii}{\mathrm{i}}

\newcommand{\calC}{\mathcal{C}}
\newcommand{\calM}{\mathcal{M}}
\newcommand{\calP}{\mathcal{P}}
\newcommand{\calT}{\mathcal{T}}
\newcommand{\calU}{\mathcal{U}}
\newcommand{\calV}{\mathcal{V}}

\newcommand{\eV}{\mathrm{eV}}
\newcommand{\keV}{\mathrm{keV}}
\newcommand{\GeV}{\mathrm{GeV}}

\renewcommand{\Im}{\mathrm{Im}}

\newcommand{\plus}{{\mbox{{\bf{\tiny +}}}}}

\begin{document}

\title{Localizability of Tachyonic Particles
and Neutrinoless Double Beta Decay}

\newcommand{\addrMST}{Department of Physics,
Missouri University of Science and Technology,
Rolla, Missouri 65409-0640, USA}

\newcommand{\addrHD}{Institut f\"{u}r Theoretische Physik,
Philosophenweg 16, 69020 Heidelberg, Germany}

\author{U. D. Jentschura}
\affiliation{\addrMST}
\affiliation{\addrHD}

\author{B. J. Wundt}
\affiliation{\addrHD}

\begin{abstract}The quantum field theory of superluminal (tachyonic) particles is
plagued with a number of problems, which include the Lorentz non-invariance of
the vacuum state, the ambiguous separation of the field operator into creation
and annihilation operators under Lorentz transformations, and the necessity of
a complex reinterpretation principle for quantum processes.  Another unsolved
question concerns the treatment of subluminal components of a tachyonic wave
packets in the field-theoretical formalism, and the calculation of the
time-ordered propagator.  After a brief discussion on related problems, we
conclude that rather painful choices have to be made in order to incorporate
tachyonic spin-$\tfrac12$ particles into field theory.  We argue that the
field theory needs to be formulated such as to allow for localizable tachyonic
particles, even if that means that a slight unitarity violation is introduced
into the $S$ matrix, and we write down field operators with unrestricted
momenta.  We find that once these choices have been made, the propagator for
the neutrino field can be given in a compact form, and the left-handedness of
the neutrino as well as the right-handedness of the antineutrino follow
naturally.  Consequences for neutrinoless double beta decay and superluminal
propagation of neutrinos are briefly discussed.
\end{abstract}


\maketitle

%
%
\section{Introduction and Overview} 

%
%
\subsection{Tachyonic Quantum Mechanics}

After the early attempts by Sudarshan {\em et al.}
(Refs.~\cite{BiDeSu1962,ArSu1968,DhSu1968,BiSu1969} and Feinberg
(Refs.~\cite{Fe1967,Fe1977}), tachyonic field theory has been scrutinized
because a number of problematic issues were discovered and discussed at length
in the literature~\cite{BrTa1968,Gl1969,Sc1971,Mu1972,
Mu1973,BaSh1974,Sh1975,%
Fe1977,Sc1982}.
Insightful reviews on the history of tachyonic quantum dynamics and tachyonic
quantum field theory are given in Refs.~\cite{Re2009,Bi2009,Bo2009}.  {\em A
priori}, Lorentz invariance does not imply that velocities greater
than the speed of light are forbidden in nature. The Lorentz transformation
singles out the speed of light as a limiting velocity, not as the maximum
velocity allowed in nature. Indeed, superluminal 
Lorentz transformations have been
discussed in some detail
in the literature (e.g., in Refs.~\cite{BiDeSu1962,SuSh1986}). 
A superluminal particle cannot be stopped,
its velocity always remains greater than the speed of light, when measured from
subluminal frames of reference.  The Einstein addition theorem remains valid
for superluminal particles.  The velocity $u'$ of a superluminal particle
measured in a moving frame, $u' = (u-v)/(1 - u \, v)$ always is superluminal,
$u' \notin (-1,1)$, if $u > 1$ is superluminal in the rest frame, and the
relative velocity of the frames fulfills $-1 < v < 1$, as a quick inspection
shows.  (We set the speed of light $c=1$ in this article.)
Tachyons are particles whose dispersion relation reads as
\begin{equation}
\label{disp}
E = \frac{m}{\sqrt{u^2 - 1}} \,,\;\;
p = \frac{m \,u}{\sqrt{u^2 - 1}} \,,\;\;
E^2 - p^2 = -m^2 \,,
\end{equation}
where $u = | \vec u | > 1$ is the 
(magnitude of the) velocity of the particle.
The ``rest frame'' of a tachyon is the frame of infinite velocity,
where the energy $E \to 0$~\cite{Fe1967,BaSh1974}.
Provided the mass term $m$ is not in itself energy-dependent,
a tachyon becomes faster as it loses energy,
by virtue of the Lorentz-invariant dispersion relation~\eqref{disp}.
Note that $p = |\vec p| = E \, u > E$, because $u > 1$.
While the relation $E^2 - p^2 = -m^2$ is obtained 
from the usual (tardyonic) dispersion relation
$E^2 - p^2 = m^2$ by the simple replacement $m \to \ii \, m$,
the equations describing tachyonic particles do 
{\em not} follow from the wave equations for subluminal 
particles by the same, simple replacement.
Rather, the following equation,
\begin{equation}
\label{grafzahl}
(\ii \gamma^\mu \, \partial_\mu - \gamma^5 \, m) \, \psi(x) = 0 \,,
\end{equation}
has been proposed for tachyonic spin-$\tfrac12$ 
particles~\cite{ChHaKo1985,Ch2000,Ch2002,JeWu2011jpa}.
Here, the $\gamma^\mu$ are the Dirac matrices, and
$\partial_\mu = \partial/\partial x^\mu$ is the derivative
with respect to space-time coordinates, whereas
$\gamma^5 = \ii \, \gamma^0 \, \gamma^1 \, \gamma^2 \, \gamma^3$
is the fifth current matrix.
In Hamiltonian form, the equation reads
\begin{equation}
\label{H5}
H_5 \, \psi(\vec r) = E \, \psi(\vec r) \,,
\qquad
H_5 = \vec\alpha \cdot \vec p + \beta \, \gamma^5 \, m \,.
\end{equation}
where $\vec \alpha = \gamma^0 \, \vec \gamma$ and 
$\beta = \gamma^0$.
This Hamiltonian is not Hermitian but 
pseudo-Hermitian~\cite{Pa1943,BeBo1998,BeDu1999,%
BeBoMe1999,BeWe2001,BeBrJo2002,
Mo2002i,Mo2002ii,Mo2002iii,Mo2003npb} and has the property
\begin{equation}
H_5(\vec r) = P \, H_5(-\vec r) \, P^{-1} = 
\calP \, H_5(\vec r) \, \calP^{-1} \,,
\end{equation}
where $P = \gamma^0$ is the matrix representation of parity and 
$\calP$ is the full parity transformation.
The pseudo-Hermitian property makes the Hamiltonian 
usable for practical calculations~\cite{BeBo1998} and 
ensures that its energy eigenvalues are 
real or come in complex-conjugate pairs.
In view of $(\beta \, \gamma^5)^2 = - \mathbbm{1}_{4 \times 4}$,
the matrix multiplying the tachyonic mass term in
Eq.~\eqref{H5} is multiplied by a matrix representation 
of the imaginary unit rather than by the imaginary unit itself.
The quantum dynamics induced by the Hamiltonian~\eqref{H5}
have been studied in Ref.~\cite{JeWu2011jpa} and imply that 
all resonance energies in the spectrum of the Hamiltonian $H_5$
fulfill the tachyonic dispersion relation~\eqref{disp}.

%
%
\subsection{Tachyonic Quantum Fields}
\label{intro_qft}

The main problems in the construction of a tachyonic 
field theory are as follows. Under a Lorentz
boost with velocity $\vec v$, as emphasized by Feinberg~\cite{Fe1967},
the energy and momentum of a particle transform as
\begin{equation}
\label{Eprime}
E' = \gamma \, (E - \vec p \cdot \vec v) \,,
\qquad
\gamma = \frac{1}{\sqrt{1-v^2}} \,,
\end{equation}
into the moving frame, in analogy to 
$t' = \gamma(t - \vec v \cdot \vec x)$
(we assume $v < 1$).
Under certain conditions, we thus transform $E > 0$
into $E' < 0$ upon Lorentz transformation, if the 
particle is tachyonic ($p > E$).
Therefore, some of the particle annihilation operators transform 
into antiparticle creation operators under Lorentz transformations; 
the separation of the field operator into creation and 
annihilation contributions is no longer Lorentz invariant. 
The sign change in $E$ occurs precisely if and only if
the time $t$ a also changes sign under a Lorentz 
transformation. Therefore, if we accept the fact
that the vacuum undergoes a Lorentz transformation, 
then the calculated cross sections actually are 
Lorentz invariant~\cite{Fe1967,Fe1977}.
This is tied to the Feinberg--Sudarshan reinterpretation
principle~\cite{Fe1967,ArSu1968}.
Note that in ordinary quantum field theory, 
the identification of antiparticles as negative-energy 
particles traveling backward in time (equivalent to 
positive-energy antiparticles traveling forward in time,
with the values of all additive physical quantities reversed)
is also tied to a reinterpretation principle.
The difference between the reinterpretation
principle for tardyonic versus tachyonic particles
is not in the reversal of the arrow of time;
it is solely in the fact that in the case of tachyonic 
particles and antiparticles, the separation into 
creation and annihilation parts is not Lorentz-invariant.

If one does not separate the field operator into 
annihilation and creation parts, then 
one has to write it down in terms of annihilation
operators only~\cite{ArSu1968,DhSu1968}. This
could seem to be attractive because the separation in this
case is Lorentz-invariant, but one then has to 
perform considerable effort in the calculation of 
a propagator, even in the case of scalar tachyonic 
particles~\cite{ArSu1968,DhSu1968}.
We therefore prefer to use the concept originally 
developed in Ref.~\cite{Fe1967,Fe1977}, even if the 
Lorentz covariance of the vacuum is extremely painful 
from the point of view of axiomatic field theory.
We note, that the Lorentz-mediated conversion 
of creation into annihilation operators is restricted
to a small kinematic domain. Using Eq.~\eqref{disp},
we have 
$E' = m (1 - u \, v)/\sqrt{u^2 - 1}$ and therefore
$E' < 0$ if and only if $u > 1/v$. For 
two laboratories on Earth traveling with respect to 
each other at a relative velocity of $v \approx 10^{-6}$,
this means that the creation/annihilation paradox affects
only those tachyonic states with $u > 10^6$, i.e.,
tachyonic particles with energies $E \lesssim m/10^3$.
For a hypothetical tachyonic neutrino with a 
tachyonic mass of $1\,\eV$, this affects neutrino 
states with energies $E \lesssim 10^{-3} \, \eV$,
which are clearly irrelevant for laboratory-based 
measurements (see also the discussion below in 
Sec.~\ref{insight} for further insight into this problem).

%
%
\subsection{Localizability}

Another conceptual problem has never been satisfactorily
addressed in the literature: namely, if we try to 
solve either the tachyonic Klein-Gordon or the 
tachyonic Dirac equation using a plane-wave
{\em ansatz} of the form $\exp(\ii \vec k \cdot \vec r)$, 
then the tachyonic energy 
\begin{equation}
E = \sqrt{\vec k^2 - m^2 - \ii \, \epsilon} 
\end{equation}
becomes imaginary for $|\vec k| > m$
(we here anticipate the $\ii \, \epsilon$ prescription
{\em \`{a} la fa\c{c}on Feynmanienne}).
If one restricts the domain of allowed momenta to 
the region $|\vec k| > m$, then the plain-wave solution
proportional to $\exp(\ii \vec k \cdot \vec r)$ do not form 
a complete set of eigenstates of the Hamiltonian any more.
Fourier transformation becomes problematic, 
and also, a tachyonic wave packet proportional to a 
Dirac-$\delta$ (eigenstate of the position operator)
cannot be expanded into plain-wave eigenstates any longer.
This has led to criticism~\cite{BrTa1968,Gl1969}.
We here argue that the states with $|\vec k| \leq m$
should {\em not} be excluded from the theory.
In the mathematical sense, the states with $|\vec k| \leq m$
constitute resonances of the tachyonic Hamiltonian
with complex resonance eigenvalues
(resonances for $\Im \, E > 0$ and antiresonances 
for $\Im \, E < 0$).
If one allows the resonances, then the domain of 
allowed wave vectors is no longer restricted. 
One can then localize (and measure the position of) a tachyonic 
particle, and furthermore, field-anticommutators 
and propagators assume their usual, localized form.
The decay (in time) of the resonances is 
thus interpreted as a genuine property of the
tachyonic field when observed from a subluminal reference frame. 
The observation of the
tachyonic wave packet from a subluminal frame
leads to the elimination
(evanescence) of subluminal components from the propagating tachyonic wave
packets; the components with 
$|\vec k| \leq m$ constitute evanescent waves.
Such a situation is not uncommon in physics.
E.g., a particle bound in a cubic anharmonic oscillator potential
is propagated via resonances with complex-valued resonance
energies~\cite{JeSuLuZJ2008}; the energies become complex
because a classical particle in a cubic potential 
escapes to infinity in finite time.
The corresponding matter waves are evanescent. 
We have no better interpretation but to assert that 
for a particle visiting us ``from the other side of the light 
barrier,'' subluminal components in wave packets must be suppressed
(evanescent) because they are incompatible with the genuine superluminal 
nature of the tachyonic particle.
The inclusion of the $\ii \epsilon$ damps the 
particle solutions (positive energy, forward in time) and the 
antiparticle solutions (negative energy, backward in time)
consistently such as to dampen the wave amplitude 
accordingly.

%
%
\subsection{Implications of the Tachyonic Formulation}

The formalism detailing all above mentioned
concepts is outlined below, for the case of 
a tachyonic spin-$\tfrac12$ particle. 
The question then is, given we are willing to invest so 
much effort in formulating a quantum field theory of tachyons, 
and given that we have to give up a number of esteemed axioms
of field theory for incorporating them, what is the return
of the investment? 
We can state that (i)~the tachyonic Dirac equation 
seems to have exactly the right symmetries for the 
description of neutrinos~\cite{JeWu2011jpa},
(ii)~the tachyonic Dirac theory
has the correct massless limit (shown below),
and (iii)~the suppression of right-handed helicity neutrino states,
and left-handed helicity antineutrino states,
follows naturally from the tachyonic theory because
the suppressed states have negative norm.
Furthermore,~(iv)~we need some form of a tachyonic 
field theory if the MINOS~\cite{AdEtAl2007}
and OPERA~\cite{OPERA2011v2} data should stand the 
test of time, and if new measurements should confirm 
low-energy neutrino data~\cite{RoEtAl1991,AsEtAl1994,AsEtAl1996,StDe1995,%
WeEtAl1999,LoEtAl1999,BeEtAl2008}
which show an apparent trend toward negative values for the 
neutrino mass square.
(The neutrino mass square exhibits a clear trend toward negative
values in all measurements, with the magnitude of the mass
square increasing with the energy of the neutrino.
The latter point is not subject of the discussion in the 
current article.)

We proceed as follows. In Sec.~\ref{solutions},
we discuss the spinor solutions of the tachyonic Dirac 
equation. Quantum field theory of tachyons is
discussed in Sec.~\ref{quant}, where we also 
calculate the propagator for tachyonic spin-$\tfrac12$ particles
based on the time-ordered product of field operators.
Neutrinoless double beta decay is discussed
in Sec.~\ref{neutrinoless}.
Finally, conclusions are reserved for Sec.~\ref{conclu}.
In particular, we here attempt to improve on an 
earlier work~\cite{BaSh1974} to quantize the tachyonic 
spin-$\tfrac12$ theory. 
The Fourier transform of our propagator is related to the
inverse of the Hamiltonian, as it should.
Units with $\hbar = c = \epsilon_0 = 1$ are used throughout 
the paper.

%
%
\section{Solutions of the Tachyonic Equation}
\label{solutions}

Starting from the solutions for the massless Dirac equation,
we proceed to the calculation of the solutions of the
tachyonic Dirac equation by an obvious generalization.
The helicity basis is a convenient starting point for the 
calculations.  The eigenfunctions of the 
operator $\vec\sigma \cdot \vec k$ are given by
\begin{subequations}
\begin{align}
a_+(\vec k) =& \; \left( \begin{array}{c} 
\cos\left(\frac{\theta}{2}\right) \\[0.77ex]
\sin\left(\frac{\theta}{2}\right) \, \ee^{\ii \, \varphi} \\
\end{array} \right) \,,
\\[0.77ex]
a_-(\vec k) =& \; \left( \begin{array}{c} 
-\sin\left(\frac{\theta}{2}\right) \, \ee^{-\ii \, \varphi} \\[0.77ex]
\cos\left(\frac{\theta}{2}\right) \\
\end{array} \right) \,.
\end{align}
\end{subequations}
Here, $\theta$ and $\varphi$ are the polar and azimuthal angles
of the wave vector $\vec k$, respectively.
They fulfill the relation
$\vec \sigma \cdot \vec k \, a_\pm(\vec k) =
\pm |\vec k|\,a_\pm(\vec k)$.
The normalized positive-energy chirality and helicity eigenstates of the 
massless Dirac equation are
\begin{equation}
\label{u}
u_+(\vec k) = 
\frac{1}{\sqrt{2}} 
\left( \begin{array}{c}
a_+(\vec k) \\[0.77ex]
a_+(\vec k) \\
\end{array} \right) \,,
\quad
u_-(\vec k) = 
\frac{1}{\sqrt{2}} 
\left( \begin{array}{c}
a_-(\vec k) \\[0.77ex]
-a_-(\vec k) \\
\end{array} \right) \,.
\end{equation}
These eigenstates immediately lead to 
plane-wave solutions of the massless Dirac equation.
Denoting by $\vec p = -\ii \vec \nabla$ the momentum operator,
the massless Dirac Hamiltonian reads $H_0 = \vec \alpha \cdot \vec p$, 
where $\vec\alpha = \gamma^0 \, \vec \gamma$ is the vector of
Dirac $\alpha$ matrices.
We use the Dirac matrices in the Dirac representation,
\begin{align}
\gamma^0 =& \; \left( \begin{array}{cc} \mathbbm{1}_{2\times 2} & 0 \\
0 & -\mathbbm{1}_{2\times 2} \\
\end{array} \right) \,,
\quad
\vec\gamma = \left( \begin{array}{cc} 0 & \vec\sigma \\ -\vec\sigma & 0  \\
\end{array} \right) \,,
\nonumber\\[0.77ex]
\gamma^5 =& \; \left( \begin{array}{cc} 0 & \mathbbm{1}_{2\times 2} \\
\mathbbm{1}_{2\times 2} & 0  \\
\end{array} \right) \,.
\end{align}
The positive-energy solutions have the properties
\begin{subequations}
\begin{align}
\vec \alpha\cdot\vec p \, u_\pm(\vec k) \, 
\ee^{\ii \vec k \cdot \vec r}
=& \; |\vec k| u_\pm(\vec k) \, 
\ee^{\ii \vec k \cdot \vec r} \,,
\\[0.77ex]
\frac{\vec\Sigma \cdot \vec p}{|\vec p|} \, u_\pm(\vec k) 
\, \ee^{\ii \vec k \cdot \vec r}
= & \; \gamma^5 \,  u_\pm(\vec k) \, \ee^{\ii \vec k \cdot \vec r}
= \pm u_\pm(\vec k) \, \ee^{\ii \vec k \cdot \vec r} \,.
\end{align}
\end{subequations}
For the negative-energy solutions of the massless Dirac
equation, we obtain the charge conjugate solutions of
the ones for positive energy,
\begin{subequations}
\label{v}
\begin{align}
v_+(\vec k) = & \; C \, \overline u_-(\vec k)^{\rm T} =
\frac{1}{\sqrt{2}} 
\left( \begin{array}{c}
-a_+(\vec k) \\[0.77ex]
-a_+(\vec k) \\
\end{array} \right) \,,
\\[0.77ex]
v_-(\vec k) =& \; C \, \overline u_+(\vec k)^{\rm T} =
\frac{1}{\sqrt{2}} 
\left( \begin{array}{c}
-a_-(\vec k) \\[0.77ex]
a_-(\vec k) \\
\end{array} \right) \,,
\end{align}
\end{subequations}
where $C = \ii \, \gamma^2 \, \gamma^0$ is the charge conjugation
matrix. The negative-energy states fulfill the relations
\begin{subequations}
\begin{align}
\vec \alpha\cdot\vec p \, v_\pm(\vec k) \, 
\ee^{-\ii \vec k \cdot \vec r}
=& \; -|\vec k| v_\pm(\vec k) \, 
\ee^{\ii \vec k \cdot \vec r} \,,
\\[0.77ex]
\label{showme}
-\frac{\vec\Sigma \cdot \vec p}{|\vec p|} \, v_\pm(\vec k) 
\, \ee^{\ii \vec k \cdot \vec r}
= & \; \gamma^5 \,  v_\pm(\vec k) \, \ee^{\ii \vec k \cdot \vec r}
= \pm v_\pm(\vec k) \, \ee^{\ii \vec k \cdot \vec r} \,.
\end{align}
\end{subequations}
The subscripts $\pm$ of the $u$ and $v$ spinors 
correspond to the chirality (eigenvalue of $\gamma^5$), 
which is equal to helicity for positive-energy eigenstates,
and equal to the negative of the chirality for 
negative-energy eigenstates.
Because of the relation
$(\gamma^\mu k_\mu - \gamma^5 \, m)^2 = k^2 + m^2 = E^2 - \vec k^{\,2} + m^2$, 
the generalization of these solutions to the massive
tachyonic Dirac equation are rather straightforward.
We find
\begin{subequations}
\label{UU}
\begin{align}
U_+(\vec k) = & \;
\frac{\gamma^5\,m- \cancel{k}}%
{\sqrt{2} \, \sqrt{(E - |\vec k|)^2 + m^2}} \, u_+(\vec k) 
\nonumber\\
=& \;
\left( \begin{array}{c}
\dfrac{m-E+|\vec k|}{\sqrt{2} \, \sqrt{(E - |\vec k|)^2 + m^2}} \; a_+(\vec k) \\[2ex]
\dfrac{m+E-|\vec k|}{\sqrt{2} \, \sqrt{(E - |\vec k|)^2 + m^2}} \; a_+(\vec k) \\
\end{array} \right) \,,
\\[2ex]
U_-(\vec k) = & \;
\frac{\cancel{k} - \gamma^5\,m}%
{\sqrt{2} \, \sqrt{(E - |\vec k|)^2 + m^2}} \, u_-(\vec k)
\nonumber\\
=& \;
\left( \begin{array}{c}
\dfrac{m+E-|\vec k|}{\sqrt{2} \, \sqrt{(E - |\vec k|)^2 + m^2}} \; a_-(\vec k) \\[2ex]
\dfrac{-m+E-|\vec k|}{\sqrt{2} \, \sqrt{(E - |\vec k|)^2 + m^2}} \; a_-(\vec k) \\
\end{array} \right) \,,
\end{align}
\end{subequations}
where $\cancel{k} = \gamma^\mu \, k_\mu$ is the Feynman dagger.
The massless limit $m \to 0$ is recovered by observing that
$E \to |\vec k|$ for a particle approaching the light cone (called 
a luxon). So, $U_+(\vec k) \to u_+(\vec k)$ and
$U_-(\vec k) \to u_-(\vec k)$ in the massless limit. 
The negative-energy eigenstates of the tachyonic Dirac 
equation are given as
\begin{subequations}
\label{VV}
\begin{align}
V_+(\vec k) = & \;
\frac{\gamma^5\,m+\cancel{k}}%
{\sqrt{2} \, \sqrt{(E - |\vec k|)^2 + m^2}} \, u_+(\vec k)
\nonumber\\
=& \;
\left( \begin{array}{c}
\dfrac{-m-E+|\vec k|}{\sqrt{2} \, \sqrt{(E - |\vec k|)^2 + m^2}} \; a_+(\vec k) \\[2ex]
\dfrac{-m+E-|\vec k|}{\sqrt{2} \, \sqrt{(E - |\vec k|)^2 + m^2}} \; a_+(\vec k) \\
\end{array} \right) \,,
\\[0.77ex]
V_-(\vec k) = & \;
\frac{-\cancel{k} - \gamma^5\,m}%
{\sqrt{2} \, \sqrt{(E - |\vec k|)^2 + m^2}} \, u_-(\vec k)
\nonumber\\
=& \;
\left( \begin{array}{c}
\dfrac{-m+E-|\vec k|}{\sqrt{2} \, \sqrt{(E - |\vec k|)^2 + m^2}} \; a_-(\vec k) \\[2ex]
\dfrac{m+E-|\vec k|}{\sqrt{2} \, \sqrt{(E - |\vec k|)^2 + m^2}} \; a_-(\vec k) \\
\end{array} \right) \,.
\end{align}
\end{subequations}
In the massless limit, as before,
we have $V_+(\vec k) \to v_+(\vec k)$ and $V_-(\vec k) \to v_-(\vec k)$. 
The states are normalized with respect to the condition
\begin{align}
U^\plus_+(\vec k) \, U_+(\vec k) =& \;
U^\plus_-(\vec k) \, U_-(\vec k) \nonumber\\[2ex]
=& \; V^\plus_+(\vec k) \, V_+(\vec k) =
V^\plus_-(\vec k) \, V_-(\vec k) = 1 \,.
\end{align}
The corresponding positive-energy solutions
of the massive tachyonic Dirac equation are given as
\begin{subequations}
\begin{align}
& \Psi(x) = U_\pm(\vec k) \, \ee^{-\ii k \cdot x} \,,
\quad
k = (E, \vec k) \,,
\quad
E = \sqrt{\vec k^{2} - m^2} \,,
\\[0.77ex]
& \left( \ii \, \gamma^\mu \, \partial_\mu - \gamma^5 m\right) \psi(x) = 
(\gamma^\mu \, k_\mu - \gamma^5 m) \Psi(x) = 0 \,,
\\[0.77ex]
& \Phi(x) = V_\pm(\vec k) \, \ee^{\ii k \cdot x} \,,
\quad
k = (E, \vec k) \,,
\quad
E = \sqrt{\vec k^2 - m^2} \,,
\\[0.77ex]
& \left( \ii \, \gamma^\mu \, \partial_\mu - \gamma^5 m\right) \psi(x) =
(-\gamma^\mu \, k_\mu - \gamma^5 m) \Phi(x) = 0 \,,
\end{align}
\end{subequations}
Here, $\Psi$ is for positive energy, and 
$\Phi$ is for negative energy.
All above formulas are valid for 
\begin{equation}
|\vec k| \geq m\,, 
\qquad
E = \sqrt{\vec k^{2} - m^2} \in \mathbbm{R} \,.
\end{equation}
For $|\vec k | < m$, one encounters resonances.
We define the width $\Gamma$ of a resonance of the tachyonic 
Dirac Hamiltonian as follows,
\begin{subequations}
\begin{align}
E =& \; \pm \sqrt{\vec k^{\,2} - m^2 - \ii \, \epsilon} = 
\mp \, \ii \, \frac{\Gamma}{2} \,,
\\[0.77ex]
\Gamma =& \; 2 \, \sqrt{m^2 - \vec k^{\,2}}  \,, 
\qquad
| \vec k | < m \,.
\end{align}
\end{subequations}
The wave functions describing the resonances are as follows,
\begin{subequations}
\label{RR}
\begin{align}
R_+(\vec k) = & \;
\left( \begin{array}{c}
\dfrac{m+\tfrac{\ii}{2} \Gamma +|\vec k|}{\sqrt{2} \, 
\sqrt{\vec k^{\,2} + m^2 + \tfrac14 \, \Gamma^2}} \; a_+(\vec k) \\[0.77ex]
\dfrac{m-\tfrac{\ii}{2} \Gamma -|\vec k|}{\sqrt{2} \, 
\sqrt{\vec k^{\,2} + m^2 + \tfrac14 \, \Gamma^2}} \; a_+(\vec k) \\
\end{array} \right) \,,
\\[0.77ex]
R_-(\vec k) = & \;
\left( \begin{array}{c}
\dfrac{m-\tfrac{\ii}{2} \Gamma -|\vec k|}{\sqrt{2} \, 
\sqrt{\vec k^{\,2} + m^2 + \tfrac14 \, \Gamma^2}} \; a_-(\vec k) \\[0.77ex]
\dfrac{-m-\tfrac{\ii}{2} \Gamma-|\vec k|}{\sqrt{2} \, 
\sqrt{\vec k^{\,2} + m^2 + \tfrac14 \, \Gamma^2}} \; a_-(\vec k) \\
\end{array} \right) \,,
\\[0.77ex]
E =& \; -\tfrac{\ii}{2}\, \Gamma = -\tfrac{\ii}{2} \, \sqrt{m - \vec k^{2}} \,,
\qquad
|\vec k| < m \,.
\end{align}
\end{subequations}
The antiresonance eigenstates are
\begin{subequations}
\label{SS}
\begin{align}
S_+(\vec k) = & \;
\left( \begin{array}{c}
\dfrac{-m-\tfrac{\ii}{2} \Gamma+|\vec k|}{\sqrt{2} \, 
\sqrt{\vec k^{\,2} + m^2 + \tfrac14 \, \Gamma^2}} \; a_+(\vec k) \\[0.77ex]
\dfrac{-m+\tfrac{\ii}{2} \Gamma-|\vec k|}{\sqrt{2} \, 
\sqrt{\vec k^{\,2} + m^2 + \tfrac14 \, \Gamma^2}} \; a_+(\vec k) \\
\end{array} \right) \,,
\\[0.77ex]
S_-(\vec k) = & \;
\left( \begin{array}{c}
\dfrac{-m+\tfrac{\ii}{2}\Gamma -|\vec k|}{\sqrt{2} \, 
\sqrt{\vec k^{\,2} + m^2 + \tfrac14 \, \Gamma^2}} \; a_-(\vec k) \\[0.77ex]
\dfrac{m+\tfrac{\ii}{2}\Gamma -|\vec k|}{\sqrt{2} \, 
\sqrt{\vec k^{\,2} + m^2 + \tfrac14 \, \Gamma^2}} \; a_-(\vec k) \\
\end{array} \right) \,,
\\[0.77ex]
E =& \; \tfrac{\ii}{2}\, \Gamma = \tfrac{\ii}{2} \, \sqrt{m - \vec k^{2}} \,,
\qquad |\vec k| < m \,.
\end{align}
\end{subequations}
These states are normalized according to 
\begin{align}
R^\plus_+(\vec k) \, R_+(\vec k) =& \;
R^\plus_-(\vec k) \, R_-(\vec k) \nonumber\\[2ex]
=& \; S^\plus_+(\vec k) \, S_+(\vec k) =
S^\plus_-(\vec k) \, S_-(\vec k) = 1  \,.
\end{align}
The rest frame for tachyons is that of infinite speed~\cite{Fe1967,BaSh1974},
which corresponds to $|\vec k| = m$ and $E \to 0$.
The eigenstates simplify in this limit to read
\begin{subequations}
\label{toUU}
\begin{align}
U_+(\vec k) \to & \;
R_+(\vec k) \to 
\left( \begin{array}{c}
a_+(\vec k) \\[0.77ex] 0 \\
\end{array} \right) \,,
\qquad |\vec k| \to m \,,
\\[0.77ex]
U_-(\vec k) \to & \;
R_-(\vec k) \to 
\left( \begin{array}{c}
0 \\[0.77ex] -a_-(\vec k) \\
\end{array} \right) \,,
\qquad |\vec k| \to m \,.
\end{align}
\end{subequations}
The negative-energy spinors tend to the following values,
\begin{subequations}
\label{toVV}
\begin{align}
V_+(\vec k) \to & \;
S_+(\vec k) \to 
\left( \begin{array}{c}
0 \\[0.77ex] -a_+(\vec k) \\
\end{array} \right) \,,
\qquad |\vec k| \to m \,,
\\[0.77ex]
V_-(\vec k) \to & \;
S_-(\vec k) \to 
\left( \begin{array}{c}
-a_-(\vec k) \\[0.77ex] 0 \\ 
\end{array} \right) \,,
\qquad |\vec k| \to m \,.
\end{align}
\end{subequations}
One observes that the
massless solutions~\eqref{u} and~\eqref{v} 
cannot be normalized according to the 
covariant expression $\overline u \, u = 
u^\plus \, \gamma^0 \, u$, because 
this expression vanishes for all four 
solutions indicated in Eqs.~\eqref{u} and~\eqref{v}.
In the massive case, we calculate after some 
algebraic simplification,
\begin{subequations}
\begin{align}
\overline U_+(\vec k) \, U_+(\vec k) = & \; \frac{m}{|\vec k|} \,, \qquad
\overline U_-(\vec k) \, U_-(\vec k) = -\frac{m}{|\vec k|} \,,
\\[0.2ex]
\overline V_+(\vec k) \, V_+(\vec k) = & \; -\frac{m}{|\vec k|} \,, \qquad
\overline V_-(\vec k) \, V_-(\vec k) = \frac{m}{|\vec k|} \,,
\end{align}
\end{subequations}
By a multiplication with $(|\vec k|/m)^{1/2}$, we can thus 
obtain covariantly normalized
spinors $\calU_\sigma(\vec k)$ and $\calV_\sigma(\vec k)$,
which fulfill the following chirality-dependent normalizations,
\begin{subequations}
\label{covariant}
\begin{align}
\overline \calU_\sigma(\vec k) \; \calU_\sigma(\vec k) =& \;
\calU^\plus_\sigma(\vec k) \gamma^0 \calU_\sigma(\vec k) = \sigma \,,
\\[2ex]
\overline \calV_\sigma(\vec k) \; \calV_\sigma(\vec k) =& \;
\calV^\plus_\sigma(\vec k) \gamma^0 \calV_\sigma(\vec k) = -\sigma \,,
\end{align}
\end{subequations}
where $\sigma = \pm$ is the chirality.
We here indicate these solutions for the eigenstates with real energy eigenvalues,
in the normalization~\eqref{covariant},
and start with the positive-energy solutions.
\begin{subequations}
\label{calUU}
\begin{align}
\calU_+(\vec k) = & \;
\left( \begin{array}{c}
\dfrac{m-E+|\vec k|}{2 \, \sqrt{m} \, \sqrt{|\vec k|- E}} \; a_+(\vec k) \\[2ex]
\dfrac{m+E-|\vec k|}{2 \, \sqrt{m} \, \sqrt{|\vec k|- E}} \; a_+(\vec k) \\
\end{array} \right) \,,
\\[0.77ex]
\calU_-(\vec k) = & \;
\left( \begin{array}{c}
\dfrac{m+E-|\vec k|}{2 \, \sqrt{m} \, \sqrt{|\vec k|- E}} \; a_-(\vec k) \\[2ex]
\dfrac{-m+E-|\vec k|}{2 \, \sqrt{m} \, \sqrt{|\vec k|- E}} \; a_-(\vec k) \\
\end{array} \right) \,.
\end{align}
\end{subequations}
The negative-energy spinors in the normalization~\eqref{covariant}
are given as
\begin{subequations}
\label{calVV}
\begin{align}
\calV_+(\vec k) = & \;
\left( \begin{array}{c}
\dfrac{-m-E+|\vec k|}{2 \, \sqrt{m} \, \sqrt{|\vec k|- E}} \; a_+(\vec k) \\[2ex]
\dfrac{-m+E-|\vec k|}{2 \, \sqrt{m} \, \sqrt{|\vec k|- E}} \; a_+(\vec k) \\
\end{array} \right) \,,
\\[0.77ex]
\calV_-(\vec k) = & \;
\left( \begin{array}{c}
\dfrac{-m+E-|\vec k|}{2 \, \sqrt{m} \, \sqrt{|\vec k|- E}} \; a_-(\vec k) \\[2ex]
\dfrac{m+E-|\vec k|}{2 \, \sqrt{m} \, \sqrt{|\vec k|- E}} \; a_-(\vec k) \\
\end{array} \right) \,.
\end{align}
\end{subequations}
In the following, we understand that for $|\vec k| < m$, 
the $\calU_\pm$ and $\calV_\pm$ are
defined as the resonances and antiresonances~\eqref{RR} and~\eqref{SS}, 
renormalized to the condition~\eqref{covariant}
in full analogy with Eqs.~\eqref{calUU} and~\eqref{calVV}.

%
%
\begin{figure}[t!]
\begin{center}
\includegraphics[width=0.8\linewidth]{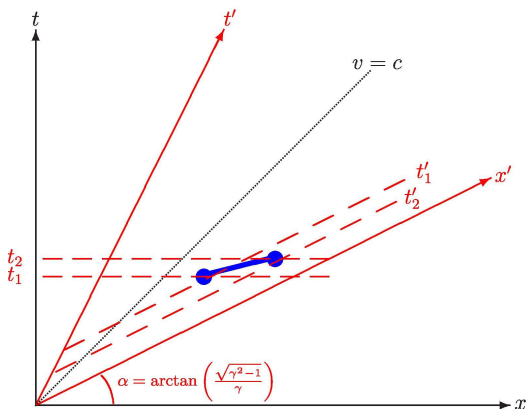}
\end{center}
\caption{\label{fig1} (Color online) 
Illustration of the Feinberg--Sudarshan 
reinterpretation principle in a Minkowski diagram:
The primed (moving) frame moves at velocity $v$
with Lorentz factor $\gamma = 1/\sqrt{1-v^2}$.
The $x'_1$ axis is tilted by the angle 
$\alpha 
= \arctan(\sqrt{1 - \gamma^{-2}})$ with 
respect to the $x_1$ axis, and 
the $t'$ axis is tilted with respect to 
$t$ by the same angle.
Two (spacelike, tachyonic) events occur at $t_2 > t_1$ and transform 
to $t'_2 < t'_1$ in the moving frame.
According to the reinterpretation principle, the 
two events are seen by the moving observer 
as a negative-energy particle annihilation event
(positive-energy antiparticle creation event) at space-time point
$(t'_1, x'_1)$. The negative-energy particle 
is created at the point $(t'_2, x'_2)$, which is equivalent to 
stating that the positive-energy antiparticle is annihilated
at $(t'_2, x'_2)$.
According to reinterpretation, an antiparticle travels 
forward in time from $(t'_2, x'_2)$ to $(t'_1, x'_1)$
in the space-time coordinates of the moving frame.
In the laboratory (rest) frame, the 
two events are seen as particle creation at 
$(t_1, x_1)$ followed by annihilation at $(t_2, x_2)$.}
\end{figure}

%
%
\section{Quantized Theory}
\label{quant}

Up to now, we have only considered the relativistic quantum theory, not the
field operator which corresponds to the tachyonic spin-$\tfrac12$ field.
According to the Feinberg--Sudarshan reinterpretation
principle~\cite{BiDeSu1962,Fe1967,ArSu1968,DhSu1968,BiSu1969}, problems
associated with the violation of causality by the emergence of tachyonic
particles can be overcome. Namely, in Ref.~\cite{ArSu1968}, the authors argue
that a physically ``sensible'' theory is achieved by insisting that the only
physical quantities are transition amplitudes and a negative-energy in (out)
state is physically understood to be a positive-energy out (in) state.  This
statement is perhaps oversimplified and in need of a further explanation.  It
can be understood as follows.  Suppose that observer $A$ sees event $E1$
before $E2$, and observer $B$ sees event $E2$ before $E1$, because the two
events are separated by a space-like interval, and the Lorentz transform
between the frames of observers $A$ and $B$ reverses the time-ordering of
events $E1$ and $E2$.  Then, according to Ref.~\cite{BiDeSu1962}, the reversal
of time ordering occurs precisely under the condition that the Lorentz
transformation between the two frames also changes the sign of the energy. So,
provided one reinterprets the negative-energy solutions of the tachyonic Dirac
Hamiltonian propagating backward in time (the antiresonances included) as
positive-energy solutions propagating forward in time, the creation and
absorption of a particle can be reinterpreted without further problems if only
the transition amplitude is unaffected by the reinterpretation.  This point
has been stressed in Refs.~\cite{BiDeSu1962,BiSu1969} and is illustrated in
Fig.~\ref{fig1}.

In Ref.~\cite{Fe1967}, both particle as well as 
antiparticle creation and annihilation operators 
are used in the field operators, and in order to 
ensure Lorentz covariance of the 
quantization conditions, anticommutators are used
in order to to quantize a scalar field
theory. The use of anticommutators is dictated 
by the fact that under Lorentz transformations, 
the energy may change sign and therefore, the commutator
of an annihilation and a creation operator 
would otherwise change sign under a Lorentz transformation,
which is inconsistent [see Eqs.~(4.8) and~(4.9) of Ref.~\cite{Fe1967}].
However, for fermions, no such problem arises, and 
we may define the quantization conditions naturally,
in terms of anticommutators.
In Refs.~\cite{ArSu1968,DhSu1968}, the authors argue that 
the field operators for the tachyonic field 
in terms of particle operators alone, which allows
the authors to quantize the scalar theory using commutators.
Concerning spin-$\tfrac12$ particles, the authors of 
Ref.~\cite{BaSh1974} write that in agreement with
Refs.~\cite{BiDeSu1962,BiSu1969},
the antiparticle operators are not included in their 
field operator, since, for tardyons, ``they
represent an interpretation of the negative-energy states; thus the inclusion
of both negative-energy states and antiparticle states is redundant.  Any
reinterpretation principle which views the negative-energy annihilation
operators as antiparticle creation operators will exclude the possibility of an
invariant vacuum state'' [see text following Eq.~(17) of Ref.~\cite{BaSh1974}].
In the formalism of~\cite{ArSu1968,DhSu1968},
antiparticle operators have to be inserted by hand into the 
tachyonic field operator, as in Eq.~(1.11) of Ref.~\cite{DhSu1968}.
{\em We explicitly accept the Lorentz-noninvariance of the 
vacuum state in the following derivation.}
According to Eq.~(5.7) of Ref.~\cite{Fe1967}, 
the Lorentz transformation of the vacuum state, in this 
case, forces us to occupy all particle and antiparticle 
states whose energy changes sign under the Lorentz
transformation, in the Lorentz-transformed vacuum state
(see also Sec.~\ref{insight} below).

In full analogy with Eq.~(3.157) of Ref.~\cite{ItZu1980}, 
and Eq.~(17) of Ref.~\cite{BaSh1974}, and in full agreement
with the Feinberg--Sudarshan reinterpretation principle, 
we thus write the field operator for the tachyonic field as
\begin{align}
\psi(x) =& \;
\int \frac{\dd^3 k}{(2\pi)^3} \, 
\frac{m}{E} \sum_{\sigma = \pm} 
\left[ b_\sigma(k) \, \calU_\sigma(\vec k) \, 
\ee^{-\ii \, k \cdot x} \right.
\nonumber\\[0.77ex]
& \; \left. 
+ b_\sigma(-k) \, \calV_\sigma(\vec k) \, 
\ee^{\ii \, k \cdot x} \right]  \,,
\nonumber\\[0.77ex]
k =& \; (E, \vec k)\,,
\qquad E = E_{\vec k} =  \sqrt{\vec k^2 - m^2 - \ii \, \epsilon} \,.
\end{align}
The second term describes the absorption of a negative-energy 
tachyonic particle that propagates backward in time 
which is equivalent to the emission of a positive-energy 
antiparticle propagating forward in time by the Feinberg-Sudarshan 
reinterpretation principle.
We thus have
\begin{align}
\label{fieldop}
\psi(x) =& \;
\int \frac{\dd^3 k}{(2\pi)^3} \, 
\frac{m}{E} \sum_{\sigma = \pm} 
\left[ b_\sigma(k) \, \calU_\sigma(\vec k) \, 
\ee^{-\ii \, k \cdot x} \right.
\nonumber\\[0.77ex]
& \; \left. 
+ d^\plus_\sigma(k) \, \calV_\sigma(\vec k) \, 
\ee^{\ii \, k \cdot x} \right]  \,,
\end{align}
where $d^\plus_\sigma$ creates antiparticles.
We postulate that the anticommutators of the 
tachyonic field operators read as 
follows (Fermi--Dirac statistics)
\begin{subequations}
\label{anticom}
\begin{align}
\left\{ a_\sigma(k) , a_{\rho}(k') \right\} = & \;
\left\{ a^\plus_\sigma(k) , a^\plus_{\rho}(k') \right\} = 0 \,,
\\[2ex]
\left\{ d_\sigma(k) , d_{\rho}(k') \right\} = & \;
\left\{ d^\plus_\sigma(k) , d^\plus_{\rho}(k') \right\} = 0 \,,
\end{align}
and the nonvanishing anticommutators are as follows,
\begin{align}
\left\{ a_\sigma(k) , a^\plus_{\rho}(k') \right\} = & \; 
(-\sigma) \, 
(2 \pi)^3 \, \frac{E}{m} \delta^3(\vec k - \vec k') \, 
\delta_{\sigma\rho}\,,
\\[2ex]
\left\{ d_\sigma(k) , d^\plus_{\rho}(k') \right\} = & \; 
(-\sigma) \, 
(2 \pi)^3 \, \frac{E}{m} \delta^3(\vec k - \vec k') \, 
\delta_{\sigma\rho}\,.
\end{align}
\end{subequations}
This implies that positive-energy states with negative chirality (and negative 
helicity in the massless limit) have to be quantized according to 
the normal Fermi--Dirac statistics, whereas the anticommutators of 
right-handed chirality field operators receive a minus sign.
We conclude that the norm of the right-handed helicity 
(positive chirality) neutrino 
one-particle state is negative, 
\begin{align}
\label{negnorm}
\left< 1_{k, \sigma} | 1_{k, \sigma} \right> =& \;
\left< 0 \left| b_\sigma(k) \, 
b^+_\sigma(k) \right| 0 \right> 
\\[0.77ex]
=& \; \left< 0 \left| \left\{ b_\sigma(k), 
b^+_\sigma(k) \right\} \right| 0 \right> 
= (-\sigma) \, V \, \frac{E}{m} \,,
\nonumber
\end{align}
where $V = (2 \pi)^3 \, \delta^3(\vec 0)$ is the normalization 
volume in coordinate space.
Thus, if the neutrino is described by the tachyonic 
Dirac equation, this implies that right-handed helicity neutrino
states have negative norm and can be excluded from the 
physical spectrum if one imposes a Gupta--Bleuler type condition
(see Chapter~3 of Ref.~\cite{ItZu1980}).
Furthermore, the norm of the positive-chirality
antineutrino states also is negative,
\begin{align}
\left< \overline 1_{k, \sigma} | \overline 1_{k, \sigma} \right> =& \;
\left< 0 \left| d_\sigma(k) \, 
d^+_\sigma(k) \right| 0 \right> 
\\[0.77ex]
=& \; \left< 0 \left| \left\{ d_\sigma(k), 
d^+_\sigma(k) \right\} \right| 0 \right> 
= (-\sigma) \, V \, \frac{E}{m} \,,
\nonumber
\end{align}
which implies that antineutrinos can only exist in 
the right-handed helicity state.
We remember that right-handed chirality 
implies left-handed helicity for antiparticles
[see Eq.~\eqref{showme}].
It is then rather easy to calculate
the matrix-values tachyonic field anticommutator, which reads as
\begin{align}
& \{ \psi(x), \overline\psi(y) \} = 
\left< 0 \left| \{ \psi(x), 
\overline\psi(y) \} \right| 0 \right> 
\nonumber\\[2ex]
& = \int \frac{\dd^3 k}{(2 \pi)^3} 
\frac{m}{2 E} \,
\sum_{\sigma = \pm} \left\{
\ee^{-\ii k \cdot (x-y)} \,
\left(-\sigma\right) \, \calU_{\sigma}(\vec k) \otimes
\overline \calU_{\sigma}(\vec k) \right.
\nonumber\\[2ex]
& \qquad \left. +
\ee^{\ii k \cdot (x-y)} \,
\left(-\sigma\right) \, \calV_{\sigma}(\vec k) \otimes
\overline \calV_{\sigma}(\vec k) \right\} \,,
\end{align}
where $\sigma$ is the chirality and $\otimes$ denotes the 
tensor product in spinor space.
The following two relations
\begin{subequations}
\label{tensor}
\begin{align}
\sum_\sigma (-\sigma) \; \calU_\sigma(\vec k) \otimes
\overline\calU_\sigma(\vec k) \,\gamma^5 =& \; 
\frac{\cancel{k} - \gamma^5 \, m}{2 m} \,,
\\[2ex]
\sum_\sigma (-\sigma) \; \calV_\sigma(\vec k) \otimes
\overline\calV_\sigma(\vec k) \,\gamma^5 =& \; 
\frac{\cancel{k} + \gamma^5 \, m}{2 m} \,,
\end{align}
\end{subequations}
are analogous to those that lead from Eq.~(3.169) to Eq.~(3.170)
of Ref.~\cite{ItZu1980}.
Note that the factor $-\sigma$ in these equations stems 
from the quantization conditions~\eqref{anticom}.
With the help of Eq.~\eqref{tensor}, we can thus 
derive the following, compact result,
\begin{equation}
\label{anticom2}
\{ \psi(x), \overline\psi(y) \} \, \gamma^5 = 
\left( \ii \, \cancel{\partial} - \gamma^5 \, m \right)
\ii \, \Delta(x - y) \,,
\end{equation}
where $\Delta(x-y)$ is the expression encountered in Eqs.~(3.55) and (3.56) 
of~\cite{ItZu1980},
\begin{equation}
\ii \, \Delta(x-y) =
\int \frac{\dd^3 k}{(2 \pi)^3} \, 
\frac{1}{2 E} \, \left[ \ee^{-\ii k \cdot (x-y)} -
\ee^{\ii k \cdot (x-y)} \right] \,,
\end{equation}
which is centered on the tachyonic mass shell.
It follows that the equal-time anticommutator reads as
\begin{align}
\label{unfiltered}
\left. \{ \psi(x), \overline\psi(y) \} \, \gamma^5 
\right|_{x_0 = y_0} 
=& \; - \gamma^0 \, 
\left. \partial_0 \, \Delta(x - y) \right|_{x_0 = y_0} 
\nonumber\\[2ex]
=& \; \gamma^0  \, \delta^3(\vec r - \vec s)
\end{align}
with the full, unfiltered Dirac-$\delta$ function
and $x =(t, \vec r)$ as well as $y= (t, \vec s)$.
This is contrast to previous ansatzes for tachyonic 
field theories~\cite{Fe1967,ArSu1968}, where the 
available momenta were restricted to the domain 
$|\vec k| \geq  m$. We here allow the resonances and 
antiresonances given in Eqs.~\eqref{RR} and~\eqref{SS}
to cover the region $|\vec k| < m$. 
Our compact result is obtained 
because the integration is over all $\vec k \in \mathbbm{R}^3$,
and the time derivative ``pulls down'' the resonance and antiresonance
energies from the arguments of the exponentials.
In passing, we note that because antiresonances
propagate backwards in time, whereas resonances
propagate forward in time, the imaginary parts of the 
resonance and antiresonance energies given in 
Eqs.~\eqref{RR} and~\eqref{SS} consistently lead to 
evanescent waves in the corresponding directions of the 
arrow of time.

Furthermore, with the help of 
Eqs.~\eqref{fieldop} and Eq.~\eqref{tensor},
after a short calculation,
one obtains the tachyonic ($T$) time-ordered 
propagator which has the representation
\begin{subequations}
\label{ST}
\begin{align}
\label{STa}
& \left< 0 \left| T \, \psi(x) \, 
\overline{\psi}(y) \gamma^5 \right| 0 \right> = 
\ii \, S_T(x - y) \,,
\\[0.77ex]
\label{STb}
& S_T(x - y) = 
\int \frac{\dd^4 k}{(2 \pi)^4} \, 
\ee^{-\ii k \cdot (x-y)} \,
\frac{\cancel{k} - \gamma^5 \, m}{k^2 + m^2 + \ii \, \epsilon} \,.
\end{align}
\end{subequations}
This is the equivalent of Eq.~(3.174) of Ref.~\cite{ItZu1980}.
As usual, the time-ordering in the time-ordered product in 
Eq.~\eqref{ST} is with regard to the field operators; the tensor 
product of the spinors is calculated according to Eq.~\eqref{tensor}.
We have thus solved the problem of the quantization 
of our tachyonic field theory, under the introduction of the 
$\gamma^5$ matrix in our time-ordered product and the normalization 
of the spinors as in Eq.~\eqref{covariant} and the 
anti-commutator relations in Eq.~\eqref{anticom}.
Because the couplings of the neutrino involve the 
chirality projector $(1 - \gamma^5)/2$, the 
$\gamma^5$ matrix in Eq.~\eqref{ST} can be absorbed in a redefinition of the 
interaction Lagrangian. Due to the equality
$\gamma^5(1 \pm \gamma^5)/2=\pm (1 \pm \gamma^5)/2$,
the chirality projectors are invariant under multiplication by $\gamma^5$.

We can otherwise explore the connection of the 
tachyonic propagator
with the Green function, in a possible analogy to the 
Dirac equation,
%
\begin{equation}
S_T = \gamma^0 \, \frac{1}{E - H_5} \,,
\end{equation}
where $E$ is the energy argument of the Green 
function and $H_5$ is the tachyonic Hamiltonian
(up to the $\ii \, \epsilon$ prescription).
An easy calculation shows that
in momentum space, with $H_5$
replaces by $\vec\alpha \cdot \vec k + \beta \gamma^5 m$,
\begin{equation}
S_T(k) = \frac{1}{\cancel{k} - \gamma^5 \, m} = 
 \frac{\cancel{k} - \gamma^5 \, m}{k^2 + m^2} \,,
\end{equation}
with $\cancel{k} = \gamma^{\mu} k_{\mu}$.
If we introduce the $\ii \epsilon$ prescription
as before, we find that 
\begin{equation}
\label{ST2}
S_T(k) = \frac{1}{\cancel{k} - \gamma^5 \, (m + \ii\,\epsilon)} =
\frac{\cancel{k} - \gamma^5 \, m}{k^2 + m^2 + \ii \, \epsilon} \,.
\end{equation}
%
This result, obtained by the inversion of the kinetic 
operator, is in full agreement with the result obtained
in Eq.~\eqref{ST} from the quantized theory.
Notably, the expression for our propagator is much more compact than that
obtained from the formalism of Ref.~\cite{BaSh1974},
where expressions become rather involved even at 
intermediate steps and the authors 
do not even proceed to the calculation  of the 
propagator [see Eq.~(20) of Ref.~\cite{BaSh1974}].

\begin{figure}[t!]
\begin{center}
\includegraphics[width=0.9\linewidth]{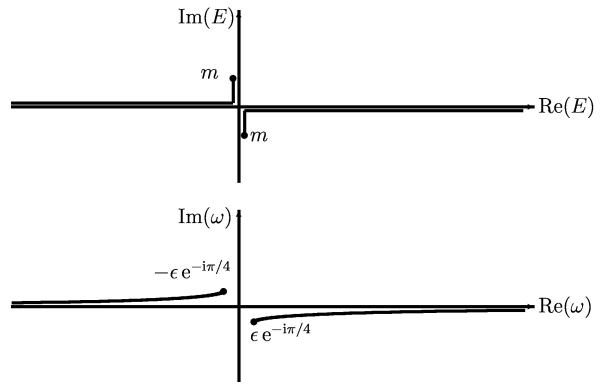}
\end{center}
\caption{\label{fig2} Branch cuts of the 
tachyonic propagator~\eqref{ST3} in the mixed 
frequency-position representation (upper panel)
and (for comparison) branch cuts of the photon propagator 
(lower panel). See text for further explanations.}
\end{figure}

The compact structure of our propagator hinges upon the 
inclusion of the resonance and antiresonance eigenstates
of the tachyonic Hamiltonian, which enables us to 
invert the kinetic operator for any $\vec k$, even 
for $|\vec k| < m$. It is instructive to consider the 
mixed space-frequency representation
\begin{equation}
\label{ST3}
S_T(E, \vec r - \vec r') = \int \frac{\dd^3 k}{(2 \pi)^3} 
S_T(E, \vec k)  \, \ee^{\ii \, \vec k \cdot (\vec r - \vec r')}  \,.
\end{equation}
For the photon propagator in the Feynman prescription, 
it is known that the branch cuts of the mixed representation
are defined to be along the lines $\sqrt{\omega^2 + \ii \, \epsilon}$
with an infinitesimal $\epsilon$,
and with the branch cut of the square root being defined to lie
along the positive real axis~\cite{Mo1974a,Mo1974b}.
For the tachyonic Hamiltonian, the branch cuts extend 
infinitesimally below the positive and above the negative real axis,
and contain small sections representing the resonances and 
antiresonances, with $| \Im(E) | \leq m$. As convenient
for loop calculations, the branch cuts of the modified tachyonic Dirac
propagator allow for a Wick rotation (see Fig.~\ref{fig2}).

%
%

%
%
\begin{figure}[t!]
\begin{center}
\includegraphics[width=0.7\linewidth]{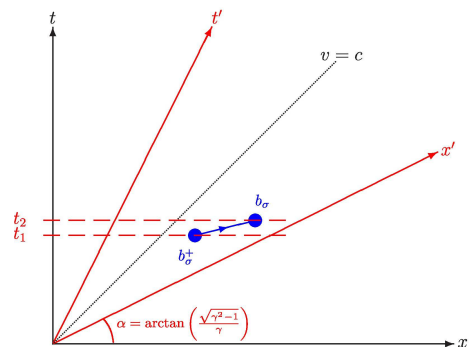}
\end{center}
\caption{\label{fig3} (Color online) Events $1$ and $2$ seen in the 
laboratory frame. A particle is created at space-time 
point $(t_1, x_1)$ and propagates to $(t_2, x_2)$ where it
is annihilated. The creation and annihilation operators 
are $b^\plus_\sigma$ and $b_\sigma$.}
\end{figure}

%
%
\begin{figure}[t!]
\begin{center}
\includegraphics[width=0.7\linewidth]{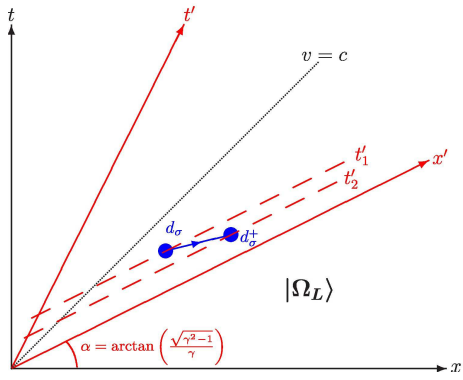}
\end{center}
\caption{\label{fig4} (Color online) Events $1$ and $2$ seen in the
moving frame, with a Lorentz-transformed vacuum 
state $| \Omega_L \rangle$. The time ordering is reversed, 
and the operator $b^\plus_\sigma$ transforms into 
an antiparticle annihilation operator $d_\sigma$,
whereas $b_\sigma$ transforms into $d^\plus_\sigma$.
In order for the matrix element to be Lorentz covariant, 
we must transform the vacuum state $|0\rangle \to 
|\Omega_L \rangle$.}
\end{figure}

%
%
\begin{figure}[t!]
\begin{center}
\includegraphics[width=0.7\linewidth]{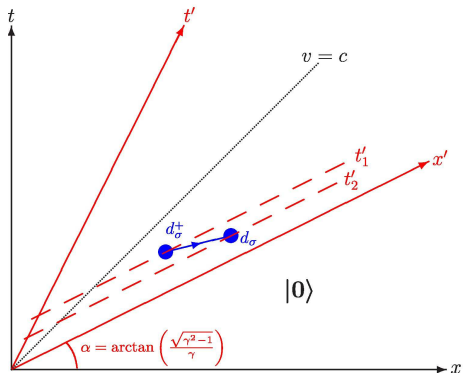}
\end{center}
\caption{\label{fig5} (Color online) The matrix element describing 
events $1$ and $2$ could also be evaluated in a different 
way by the moving observer. The trajectory is backward in time.
Therefore, it is an antiparticle trajectory in the 
moving frame. 
Using the untransformed vacuum, we evaluate,
in the moving frame,
the transition amplitude by the appropriate
antiparticle creation and annihilation operators.
This reverses the ordering of the annihilation 
and creation operators with respect to Fig.~\ref{fig4}
and leads to the result given in Eq.~\eqref{hurrrahhh}.}
\end{figure}

%
%
\section{Insight into Lorentz Invariance}
\label{insight}


Let us consider the process shown in Fig.~\ref{fig1} 
in closer detail and choose the arrow of time
as indicated in Fig.~\ref{fig2}. 
In the laboratory frame, the 
particle moves from space-time point 
$(t_1, x_1)$ to space-time point $(t_2, x_2)$.
We recall the form of the field operator
\begin{align}
\psi(x) =& \;
\int \frac{\dd^3 k}{(2\pi)^3} \, 
\frac{m}{E} \sum_{\sigma = \pm} 
\left[ b_\sigma(k) \, \calU_\sigma(\vec k) \, 
\ee^{-\ii \, k \cdot x} \right.
\nonumber\\[0.77ex]
& \; \left. 
+ d^\plus_\sigma(k) \, \calV_\sigma(\vec k) \, 
\ee^{\ii \, k \cdot x} \right]  \,,
\end{align}
whose Dirac adjoint is 
\begin{align}
\overline\psi(x) =& \;
\int \frac{\dd^3 k}{(2\pi)^3} \, 
\frac{m}{E} \sum_{\sigma = \pm} 
\left[ b^\plus_\sigma(k) \, \overline\calU_\sigma(\vec k) \, 
\ee^{\ii \, k \cdot x} \right.
\nonumber\\[0.77ex]
& \; \left. 
+ d_\sigma(k) \, \overline\calV_\sigma(\vec k) \, 
\ee^{-\ii \, k \cdot x} \right]  \,.
\end{align}
In momentum space, according to Eq.~\eqref{anticom},
the amplitude for creation of the 
particle at $(t_1, x_1)$ and annihilation of the 
particle at $(t_2, x_2)$ is proportional to the 
matrix element (vacuum expectation value)
\begin{align}
\calM =& \; \left< 0 \left| b_\sigma(k_1) b^\plus_\sigma(k_2)
\right| 0 \right> = 
\{ b_\sigma(k_1), b^\plus_\sigma(k_2) \}
\nonumber\\[2ex]
=& \; (-\sigma) \, 
(2 \pi)^3 \, \frac{E}{m} \delta^3(\vec k_1 - \vec k_2) \,.
\end{align}
The Lorentz transform of this expression is
\begin{align}
\calM' =& \; L \, \calM \, L^{-1} =
\{ L b_\sigma(k_1) L^{-1}, L \, b^\plus_\sigma(k_2) L^{-1} \}
\nonumber\\[2ex]
=& \; (-\sigma) \, 
(2 \pi)^3 \, \frac{E'}{m} \delta^3(\vec k'_1 - \vec k'_2) \,.
\end{align}
where $L$ is the representation of the Lorentz transformation
in the space of the operators,
and $k'_1 = (E'_1, \vec k'_1)$ is the Lorentz transform
of $k_1 = (E_1, \vec k_1)$. 
According to Eq.~\eqref{Eprime},
$E'_1$ can be negative if $E_1$ is positive,
and we consider this situation in the Minkowski 
diagrams given in Figs.~\ref{fig3}---\ref{fig5}.
According to Fig.~\ref{fig4}, the time ordering and
the sign of the energy is 
reversed in the primed (moving frame). 
The Lorentz transform of the 
corresponding particle annihilation 
operator is thus calculated,
according to Eq.~(4.7) of Ref.~\cite{Fe1967}, as follows,
\begin{equation}
L \, b_\sigma(k_1) \, L^{-1} = b_\sigma(-k'_1) 
= d_\sigma^\plus(k'_1) \,,
\end{equation}
i.e.~the particle annihilation operator turns 
into an antiparticle creation operator.
We can easily show that the anticommutator is Lorentz-covariant,
\begin{align}
\calM' =& \{ d^\plus_\sigma(k'_1), d^\plus_\sigma(k'_2) \}
\nonumber\\[2ex]
=& \; (-\sigma) \, 
(2 \pi)^3 \, \frac{E'}{m} \delta^3(\vec k'_1 - \vec k'_2) 
\end{align}
but we cannot use the vacuum expectation value because
\begin{equation}
\left< 0 \left| L b_\sigma(k_1) L^{-1} 
L b^\plus_\sigma(k_2) L^{-1} \right| 0 \right> =
\left< 0 \left| d^\plus_\sigma(k_1) 
d_\sigma(k_2)  \right| 0 \right> = 0 \,.
\end{equation}
The vacuum cannot be Lorentz-invariant.
We have to transform it according to 
$| \Omega_L \rangle = L \, |0 \rangle$,
so that 
\begin{align}
\label{MM}
\calM' =& \; \left< \Omega_L \left| 
d^\plus_\sigma(k'_1) \,
d_\sigma(k'_2) \, \right| \Omega_L \right>  
\nonumber\\[2ex]
=& \; (-\sigma) \, 
(2 \pi)^3 \, \frac{E'}{m} \delta^3(\vec k'_1 - \vec k'_2) 
\end{align}
holds. This can be achieved as follows.
Namely, according to Eq.~(5.7) of Ref.~\cite{Fe1967},
the Lorentz-transformed vacuum is actually filled 
with all particle and antiparticle states whose energy 
changes sign under the Lorentz transformation. Therefore,
\begin{align}
\calM' =& \; 
\int \frac{\dd^3 k'}{(2 \pi)^3}  \, \frac{m}{E'} 
(-\rho) 
\left< 0 \left| 
d_\rho(k') \, d^\plus_\sigma(k'_1) \,
d_\sigma(k'_2) \, d_\rho^\plus(k') \right| 0 \right>  
\nonumber\\[2ex]
=& \; 
\int \frac{\dd^3 k'}{(2 \pi)^3}  \, \frac{m}{E'} 
(-\rho) \,
\nonumber\\[2ex]
& \; \times \left< 0 \left| 
\{ d_\rho(k'), d^\plus_\sigma(k'_1) \} \,
\{ d_\sigma(k'_2), d_\rho^\plus(k') \} \right| 0 \right>  
\nonumber\\[2ex]
=& \; 
\int \frac{\dd^3 k'}{(2 \pi)^3}  \, \frac{m}{E'}  \,
(-\rho) \, (-\sigma) \,  \delta_{\sigma \rho}
(2 \pi)^3 \, \frac{E'}{m} \delta^3(\vec k' - \vec k'_1)
\nonumber\\[2ex]
& \; \times (-\sigma) \,  \delta_{\sigma \rho}
(2 \pi)^3 \, \frac{E'}{m} \delta^3(\vec k' - \vec k'_2)
\nonumber\\[2ex]
=& \; (-\sigma) \,
(2 \pi)^3 \, \frac{E'}{m} \delta^3(\vec k'_1 - \vec k'_2)
\end{align}
where the integral over $\dd^3k'$ is over those 
states whose energy changes sign and
the factor $(-\rho)$ in the Lorentz-transformed 
vacuum ensures that it has positive norm.
(For the sum over the
chiralities $\rho$ of the occupied states,
we appeal to the summation convention.)
This confirms that upon acting on the field operators
and on the vacuum with the proper Lorentz
transformation, the matrix $\calM$ is covariant,
but the necessity for a Lorentz transformation
of the vacuum is unsettling.
Furthermore,
according to Eq.~\eqref{MM}, in the Lorentz-transformed 
expression,
the creation operator stands before the annihilation
operator, which is somewhat unnatural.

Suppose, though, that the moving observer had
used a Lorentz-invariant vacuum $|0\rangle$ and 
had evaluated the process according to 
form-invariant field operators which differ from those
of the laboratory frame only in the space-time arguments.
According to Fig.~\ref{fig5}, the moving observer
identifies antiparticle creation with momentum 
$k'_1$ and antiparticle annihilation with 
momentum $k'_2$. So, the amplitude evaluated by the 
moving observer using shape-invariant 
field operators and a Lorentz-invariant vacuum is 
\begin{equation}
\label{hurrrahhh}
\calM' = \left< 0 \left| d_\sigma(k'_2) \,
d^\plus_\sigma(k'_1) \right| 0 \right>  
= (-\sigma) \,
(2 \pi)^3 \, \frac{E'}{m} \, \delta^3(\vec k'_1 - \vec k'_2) \,,
\end{equation}
which is exactly equal to the result obtained 
using the Lorentz-transformed vacuum, and the 
Lorentz-transformed field operators.

We here conjecture that a generalization of this 
statement should hold, namely, that the 
Lorentz-transformed probability amplitude for 
any process, calculated using shape-invariant 
field operators for the tachyonic field (where
creators and annihilators retain their entity
and only the space-time arguments are transformed)
is equal to the amplitude calculated using 
Lorentz-transformed field operators (which mix 
creators and annihilators) and the Lorentz-transformed 
vacuum. In this context, it is instructive to 
observe that the form of the tachyonic propagator 
given in Eq.~\eqref{ST} involves only Lorentz-invariant 
quantities.

%
%
\section{Neutrinoless Double Beta Decay}
\label{neutrinoless}

Finally, let us briefly comment on charge conjugation.
It is known and relatively easy to show that the (ordinary,
tardyonic) Dirac equation is invariant under charge 
conjugation, provided one changes the sign of the 
charge $e \to -e$ when describing the antiparticle.
For neutral particles, the Dirac equation thus is 
fully charge conjugation invariant. 
Therefore, it is possible to construct Majorana 
solutions which are invariant under charge conjugation.
For the massless Dirac equation, we have
\begin{equation}
\Psi_+(x) = 
u_{(+)} (\vec k) \ee^{- \ii k \cdot x} + 
(-v_-(\vec k)) \ee^{\ii k \cdot x} \,,
\end{equation}
which is invariant because
\begin{align}
\Psi^{\cal C}_+(x) =& \; C \bar{\Psi}_+(x)^{\rm T}
\nonumber\\
=& \; C \, \overline{u}_+(\vec k)^{\rm T} \ee^{\ii k \cdot x} + 
C \, (-\overline{v}_-(\vec k)^{\rm T}) \ee^{-\ii k \cdot x} 
\nonumber\\
=& \; (-v_{-}(\vec k)) \ee^{\ii k \cdot x} + 
u_{+} (\vec k) \ee^{-\ii k \cdot x} 
=  \Psi_+(x) \,.
\end{align}
However, the tachyonic Dirac equation is not charge conjugation
invariant but changes~\cite{JeWu2011jpa} under charge conjugation to
\begin{equation}
\label{grafzahlC}
(\ii \gamma^\mu \, \partial_\mu + \gamma^5 \, m) \psi^{\calC}(x) = 0 \,,
\end{equation}
i.e., the sign of the $\gamma^5$ term reverses its sign.
This is because the 
tachyonic Dirac equation is $\calC\calP$,  and $\calT$ invariant,
but not $\calC$ invariant~\cite{JeWu2011jpa}. It thus describes a 
tachyonic fermionic field with particle and antiparticle states
which are manifestly different.
It is rather straightforward to calculate the
properties
\begin{subequations}
\begin{align}
\calU^{\cal C} _+(\vec k) =& \;
C \overline \calU_+(\vec k)^{\rm T} = \calU_-(\vec k) \,,
\\[0.2ex]
\calU^{\cal C} _-(\vec k) =& \;
C \overline \calU_-(\vec k)^{\rm T} = \calU_+(\vec k) \,,
\\[0.2ex]
\calV^{\cal C} _+(\vec k) =& \;
C \overline \calV_+(\vec k)^{\rm T} = \calV_-(\vec k) \,,
\\[0.2ex]
\calV^{\cal C} _-(\vec k) =& \;
C \overline \calV_-(\vec k)^{\rm T} = \calV_+(\vec k) \,,
\end{align}
\end{subequations}
which hold for the solutions of the tachyonic Dirac equation.
Particle solutions are not transferred to antiparticle 
solutions by charge conjugation, but rather, to 
solutions with the opposite chirality,
in full agreement with the relation
of Eq.~\eqref{grafzahl} to~\eqref{grafzahlC}
[$\gamma^5 \to -\gamma^5$].
It thus seems unlikely that one could modify the 
tachyonic Dirac equation such as to allow for 
charge conjugation invariant solutions, 
or, to construct a Majorana field from the solutions
of the tachyonic Dirac equation.
Neutrinoless double decay thus is not allowed
if we assume that the neutrino is described by the 
tachyonic Dirac equation.

%
%
\section{Conclusions}
\label{conclu}

Let us summarize a few observations made in our investigations.
Feinberg~\cite{Fe1967} noted that if one writes the field operator with
creation and annihilation operators, then it is impossible to impose
anticommutation relations for the field operators, because under a Lorentz
transformation, creation operators may transform into annihilation operators
and vice versa, which reverses the sign of the quantization condition if
commutators are used [see Eq.(4.8) of Ref.~\cite{Fe1967}].  He therefore
suggested to quantize a scalar, tachyonic theory using anticommutators,
effectively imposing Fermi-Dirac statistics onto scalar particles.
Alternatively, we may interpret the necessity to invoke anticommutators
instead of commutators for the quantization of a tachyonic theory as
suggesting that only fermions are suitable candidates for tachyonic particles.
{\em This observation applies to the neutrinos which are 
spin-$\tfrac12$ particles.}

We here introduce the tachyonic Dirac equation [Eq.~\eqref{grafzahl}] which is
obtained from the ordinary (tardyonic) Dirac equation by a matrix-valued
representation of the imaginary unit $\ii \to \beta \, \gamma^5$.  The spinor
solutions to the equation have some peculiar properties and normalizations and
imply special anticommutation relations~\eqref{anticom} which imply that the
right-handed neutrino states have negative norm
[see Eq.~\eqref{negnorm}].  Such states,
for the indefinite-metric photon field, are excluded from the physical states
by a Gupta-Bleuler type condition while being present in the propagator. 
{\em This observation applies to the neutrinos which have never been
observed in right-handed helicity states.}

The tachyonic Dirac equation implies to particles 
which cannot be stopped; they remain 
superluminal in any subluminal reference frame. 
The rest frame is that of infinite velocity
[see Eqs.~\eqref{toUU} and~\eqref{toVV}].
The charge conjugate of the tachyonic Dirac equation 
implies that the antiparticles of particles
described by the tachyonic Dirac equation differ
only in the sign of the chirality, or helicity
[cf.~Eqs.~\eqref{grafzahl} and~\eqref{grafzahlC}].
Furthermore, the tachyonic Dirac equation is
$\calC \calP$ invariant, where we observe that the 
parity transformation again restores the original sign 
of the chirality. 
{\em This observation applies 
to neutrinos and antineutrinos which differ in the 
sign of the helicity and have never been observed at rest.}

We have therefore carried out a more detailed 
analysis of the spinor solutions of the tachyonic 
Dirac equation (Sec.~\ref{solutions}),
before quantizing the theory in Sec.~\ref{quant},
using anticommutators. Further considerations
regarding the problematic Lorentz covariance of the 
vacuum state and possible solutions of this problem 
are described in Sec.~\ref{insight}.
Finally, in Sec.~\ref{neutrinoless} we conclude that 
if the neutrino is a tachyonic Dirac particle, 
neutrinoless double beta decay is forbidden.

A few remarks on the experimental status are in order.
{\em All of the following remarks are somewhat 
speculative at the current time and should thus be taken with 
a grain of salt.}
The general trend is measurements of the 
neutrino mass square points to negative (tachyonic) values.
By inspection of neutrino 
data~\cite{DaEtAl1987,RoEtAl1991,AsEtAl1994,StDe1995,AsEtAl1996,
WeEtAl1999,LoEtAl1999,AdEtAl2007,BeEtAl2008,OPERA2011v2}
(for a summary overview see~\cite{LABneutrino}),
one may conclude that neutrinos at higher energy 
(in the $\GeV$~range)
exhibit a larger tachyonic mass square than
those observed at low energies
(in the $\keV$~range).
The trend of all data is toward negative values of the 
mass square. If the trend is the data is confirmed,
it implies a ``running'' of the effective 
neutrino mass with the energy, which we shall not 
discuss in any further detail here.
(The neutrino mass running might otherwise suppress
conceivable decay processes of the Cerenkov type,
because the neutrino mass in the final state 
of the decay process may substantially differ from the 
effective neutrino mass in the initial state of the 
decay process.)

The trend in the observed neutrino 
masses~\cite{DaEtAl1987,RoEtAl1991,AsEtAl1994,StDe1995,AsEtAl1996,
WeEtAl1999,LoEtAl1999,AdEtAl2007,BeEtAl2008,OPERA2011v2}
implies that the neutrino approaches 
the light cone closer and closer as its energy 
decreases, because its effective 
tachyonic mass tends to small values (in the 
$\eV$ range) for a total neutrino energy in the 
$\keV$ range~\cite{RoEtAl1991,AsEtAl1994,StDe1995,AsEtAl1996}. 
Even in the $\GeV$ range, the deviation toward
superluminal velocities is ``only'' in the range
of a few parts in $10^{5}$
(see Refs.~\cite{AdEtAl2007,OPERA2011v2}).
The conceivable violation of the causality 
principle implied 
by the small deviation of the propagation velocity 
toward superluminal velocities is thus restricted to 
small kinematic regions, but still 
causes a number of fundamental concerns. 
Tachyonic theory, including the reinterpretation 
principle discussed in the current paper, 
may solve a number of these issues.
One may speculate about possible restrictions on ``allowed''
violations of the causality principle observed in 
subluminal reference frames. For instance,
one might conceive that 
violations of the causality principle could be subject 
a generalized ``uncertainty principle'' which describes the allowed
deviations from the velocity of light for particles 
in a specific energy, or frequency range.

Sudarshan is being quoted in~\cite{Re2009} with
reference to an imaginary demographer who studies population patterns on
the Indian subcontinent:
``Suppose a demographer calmly asserts that there are no people North of the
Himalayas, since none could climb over the mountain ranges! That would be an
absurd conclusion. People of central Asia are born there and live there: they
did not have to be born in India and cross the mountain range. So with
faster-than-light particles.''
If neutrinos are faster-than-light particles, then, 
since the deviations from $c$ are small~\cite{AdEtAl2007,OPERA2011v2}, 
it might well be possible that, figuratively speaking, we are allowed to 
``summit'' and to take a glance over the top of the 
mountain range, but not much more.
Furthermore, if we try to do so over longer time intervals,
i.e., at smaller energies, then the decrease of the effective
neutrino mass with the energy implies that we are less and less
allowed to do. Further considerations on these issues
are beyond the scope of the current article.
{\em Again, the statements in the last three 
paragraphs above are somewhat speculative and should 
be taken with a grain of salt; they still seem to be 
in order in view of a topical experimental result~\cite{OPERA2011v2}.}

{\em Note added.} 
The tachyonic Dirac equation naturally implies a
physically different behaviour of the chirality components of the 
neutrino field. The same observation has 
been made, based on different arguments, in the 
recent preprint by Dartora and Cabrera, 
{\em The electroweak theory with a priori superluminal 
neutrinos and its physical consequences}, arXiv:~1112.3050v1.

Finally, a technical note:
The presence of the $\gamma^5$ matrix in 
the time-ordered vacuum expectation value in Eq.~\eqref{STa} 
can otherwise be understood as follows. 
With the $\ii \epsilon$ prescription, the energy-momentum 
representation of the tachyonic propagator in Eq.~\eqref{STb} reads
\begin{equation}
S_T(k) = \frac{1}{\cancel{k} - \gamma^5 \, m\,(1 + \ii \, \epsilon)} \,.
\end{equation}
This implies a compact energy-momentum space representation for 
the vacuum expectation value of 
the time-ordered product (without $\gamma^5$):
\begin{equation}
\left< 0 \left| T \, \psi(x) \, 
\overline{\psi}(y) \right| 0 \right> = 
\int \frac{\dd^4 k}{(2 \pi)^4} \, 
\frac{\ee^{-\ii k \cdot (x-y)}}%
{\gamma^5 \, \cancel{k} - m\,(1 + \ii \, \epsilon)} \,,
\end{equation}
where we observe that the integrand simplifies to $1/(\gamma^5 \, \cancel{k} - m)$
without the $\ii \, \epsilon$ prescription.

%
%
\section*{Acknowledgments}

This work was supported by the NSF and by the
National Institute of Standards and Technology
(precision measurement grant).
One of the authors (U.D.J.) is grateful for kind hospitality at the 
Universidad de San Nicolas de Hidalgo
in Morelia (Michoacan, Mexico), where part of this 
work was completed.


\begin{thebibliography}{47}

\bibitem{BiDeSu1962}
O.M.P. Bilaniuk, V.K. Deshpande, E.C.G. Sudarshan, Am. J. Phys. \textbf{30},
  718 (1962)

\bibitem{ArSu1968}
M.E. Arons, E.C.G. Saudarshan, Phys. Rev. \textbf{173}, 1622 (1968)

\bibitem{DhSu1968}
J.~Dhar, E.C.G. Saudarshan, Phys. Rev. \textbf{174}, 1808 (1968)

\bibitem{BiSu1969}
O.M. Bilaniuk, E.C.G. Sudarshan, Nature (London) \textbf{223}, 386 (1969)

\bibitem{Fe1967}
G.~Feinberg, Phys. Rev. \textbf{159}, 1089 (1967)

\bibitem{Fe1977}
G.~Feinberg, Phys. Rev. D \textbf{17}, 1651 (1978)

\bibitem{BrTa1968}
M.M. Broido, J.G. Taylor, Phys. Rev. \textbf{174}, 1606 (1968)

\bibitem{Gl1969}
M.~Gl\"{u}ck, Phys. Rev. \textbf{183}, 1514 (1969)

\bibitem{Sc1971}
B.~Schroer, Phys. Rev. D \textbf{3}, 1764 (1971)

\bibitem{Mu1972}
J.E. Murphy, Phys. Rev. D \textbf{6}, 426 (1972)

\bibitem{Mu1973}
J.E. Murphy, Phys. Rev. D \textbf{7}, 1260 (1973)

\bibitem{BaSh1974}
J.~Bandukwala, D.~Shay, Phys. Rev. D \textbf{9}, 889 (1974)

\bibitem{Sh1975}
D.~Shay, J. Math. Phys. \textbf{16}, 1934 (1975)

\bibitem{Sc1982}
C.~Schwartz, Phys. Rev. D \textbf{25}, 356 (1982)

\bibitem{Re2009}
E.~Recami, J. Phys. Conf. Ser. \textbf{196}, 012020 (2009)

\bibitem{Bi2009}
O.M. Bilaniuk, J. Phys. Conf. Ser. \textbf{196}, 012021 (2009)

\bibitem{Bo2009}
S.K. Bose, J. Phys. Conf. Ser. \textbf{196}, 012022 (2009)

\bibitem{SuSh1986}
R.I. Sutherland, J.R. Shepanski, Phys. Rev. D \textbf{33}, 2896 (1986)

\bibitem{ChHaKo1985}
A.~Chodos, A.I. Hauser, V.A. Kostelecky, Phys. Lett. B \textbf{150}, 431 (1985)

\bibitem{Ch2000}
T.~Chang, {\em A new Dirac-type equation for tachyonic neutrinos}, e-print
  hep-th/0011087

\bibitem{Ch2002}
T.~Chang, Nucl. Sci. Technol. \textbf{13}, 129 (2002)

\bibitem{JeWu2011jpa}
U.D. Jentschura, B.J. Wundt, {\em Symmetries of the Tachyonic Dirac Equation},
  submitted, a preliminary version can be found at e-print arXiv:1110.4171
  [hep-ph]

\bibitem{Pa1943}
W.~Pauli, Rev. Mod. Phys. \textbf{15}, 175 (1943)

\bibitem{BeBo1998}
C.M. Bender, S.~Boettcher, Phys. Rev. Lett. \textbf{80}, 5243 (1998)

\bibitem{BeDu1999}
C.M. Bender, G.V. Dunne, J. Math. Phys. \textbf{40}, 4616 (1999)

\bibitem{BeBoMe1999}
C.M. Bender, S.~Boettcher, P.N. Meisinger, J. Math. Phys. \textbf{40}, 2201
  (1999)

\bibitem{BeWe2001}
C.M. Bender, E.J. Weniger, J. Math. Phys. \textbf{42}, 2167 (2001)

\bibitem{BeBrJo2002}
C.M. Bender, D.C. Brody, H.F. Jones, Phys. Rev. Lett. \textbf{89}, 270401
  (2002)

\bibitem{Mo2002i}
A.~Mostafazadeh, J. Math. Phys. \textbf{43}, 205 (2002)

\bibitem{Mo2002ii}
A.~Mostafazadeh, J. Math. Phys. \textbf{43}, 2814 (2002)

\bibitem{Mo2002iii}
A.~Mostafazadeh, J. Math. Phys. \textbf{43}, 3944 (2002)

\bibitem{Mo2003npb}
A.~Mostafazadeh, J. Math. Phys. \textbf{44}, 974 (2003)

\bibitem{JeSuLuZJ2008}
U.D. Jentschura, A.~Surzhykov, M.~Lubasch, J.~Zinn-Justin, J. Phys. A
  \textbf{41}, 095302 (2008)

\bibitem{AdEtAl2007}
P.~Adamson, C.~Andreopoulos, K.E. Arms, R.~Armstrong, D.J. Auty, S.~Avvakumov,
  D.S. Ayres, B.~Baller, B.~Barish, P.D. Barnes et~al., Phys. Rev. D
  \textbf{76}, 072005 (2007)

\bibitem{OPERA2011v2}
T. Adam {\em et al.}, OPERA Collaboration, {\em Measurement of the neutrino
  velocity with the OPERA detector in the CNGS beam}, e-print arXiv:1109.4897v2

\bibitem{RoEtAl1991}
R.G.H. Robertson, T.J. Bowles, G.J. Stephenson, D.L. Wark, J.F. Wilkerson, D.A.
  Knapp, Phys. Rev. Lett. \textbf{67}, 957 (1991)

\bibitem{AsEtAl1994}
K.~Assamagan, C.~Br\"{o}nnimann, M.~Daum, H.~Forrer, R.~Frosch, P.~Gheno,
  R.~Horisberger, M.~Janousch, P.R. Kettle, T.~Spirig et~al., Phys. Lett. B
  \textbf{335}, 231 (1994)

\bibitem{AsEtAl1996}
K.~Assamagan, C.~Br\"{o}nnimann, M.~Daum, H.~Forrer, R.~Frosch, P.~Gheno,
  R.~Horisberger, M.~Janousch, P.R. Kettle, T.~Spirig et~al., Phys. Rev. D
  \textbf{53}, 6065 (1996)

\bibitem{StDe1995}
W.~Stoeffl, D.J. Decman, Phys. Rev. Lett. \textbf{75}, 3237 (1995)

\bibitem{WeEtAl1999}
C.~Weinheimer, B.~Degen, A.~Bleile, J.~Bonn, L.~Bornschein, O.~Kazachenko,
  A.~Kovalik, E.~Otten, Phys. Lett. B \textbf{460}, 219 (1999)

\bibitem{LoEtAl1999}
V.M. Lobashev, V.N. Aseev, A.I. Belesev, A.I. Berlev, E.V. Geraskin, A.A.
  Golubev, O.V. Kazachenko, Y.E. Kuznetsov, R.P. Ostroumov, L.A. Ryvkis et~al.,
  Phys. Lett. B \textbf{460}, 227 (1999)

\bibitem{BeEtAl2008}
A.I. Belesev, E.V. Geraskin, B.L. Zhuikov, S.V. Zadorozhny, O.V. Kazachenko,
  V.M. Kohanuk, N.A. Lihovid, V.M. Lobasheva, A.A. Nozik, V.I. Parfenov et~al.,
  Phys. At. Nucl. \textbf{71}, 449 (2008)

\bibitem{ItZu1980}
C.~Itzykson, J.B. Zuber, \emph{\relax{Quantum Field Theory}} (McGraw-Hill, New
  York, 1980)

\bibitem{Mo1974a}
P.J. Mohr, Ann. Phys. (N.Y.) \textbf{88}, 26 (1974)

\bibitem{Mo1974b}
P.J. Mohr, Ann. Phys. (N.Y.) \textbf{88}, 52 (1974)

\bibitem{DaEtAl1987}
V.L. Dadykin, G.T. Zatsepin, V.B. Karchagin, P.V. Korchagin, S.A. Mal'gin, O.G.
  Ryazhskaya, V.G. Ryasnyi, V.P. Talochkin, F.F. Khalchukov, V.F. Yakushev
  et~al., JETP \textbf{45}, 593 (1987)

\bibitem{LABneutrino}
see the URL http://cupp.oulu.fi/neutrino/nd-mass.html

\end{thebibliography}
\end{document}